\documentclass{PoS}
\usepackage{epsfig,graphicx,wrapfig,amsmath,amssymb}
\newcommand{\be}{\begin{equation}}
\newcommand{\ee}{\end{equation}}
\newcommand{\ba}{\begin{eqnarray}}
\newcommand{\ea}{\end{eqnarray}}
\newcommand{\di}{\!{\rm d}}
\newcommand{\la}{\langle}
\newcommand{\ra}{\rangle}
\newcommand{\Pnew}{P_{\!\Delta c}}

\title{Structure of the Energy-Momentum Tensor and Applications}
\ShortTitle{Structure of Energy-Momentum Tensor and Applications}
	\author{Jonathan Hudson,\\ 
	Department of Physics, University of Connecticut, Storrs, CT 06269, USA}
	\author{Irina A.~Perevalova,\\
	Physics Department, Irkutsk State University, 
	Karl Marx str.~1, 664003, Irkutsk, Russia}
	\author{Maxim V.~Polyakov,\\
	Petersburg Nuclear Physics Institute, 
	Gatchina, 188300, St.~Petersburg, Russia\\
	Institut f\"ur Theoretische Physik II, 
	Ruhr-Universit\"at Bochum, D-44780 Bochum, Germany}
	\author{\speaker{Peter Schweitzer}\\
	Department of Physics, University of Connecticut, Storrs, CT 06269, USA
	\\ E-mail: \email{peter.schweitzer@phys.uconn.edu}}

\abstract{
The probably most fundamental information about a particle is contained in 
the matrix elements of its energy momentum tensor (EMT) which are accessible 
from hard-exclusive reactions via generalized parton distribution functions. 
The spin decomposition of the nucleon and Ji sum rule are one example. 
Less prominent but equally important information is encoded in the 
stress tensor, related to the spatial components of the EMT, which shows in 
detail how the strong forces inside the nucleon balance to form a bound state. 
This provides not only unique insights on nucleon structure. It also leads 
to fascinating new applications to hadron spectroscopy which allow us to 
formulate new interpretations of the charmonium-nucleon pentaquarks discovered 
by LHCb. Recent progress is reviewed in this short overview article.}

\FullConference{QCD Evolution 2016\\
		May 30-June 03, 2016\\
		National Institute for Subatomic Physics (Nikhef), Amsterdam}

\begin{document}

\section{Introduction}

The matrix elements of the EMT define some of the most fundamental properties 
of a particle: mass, spin, and $D$-term. The first two are well-known 
properties, but the latter is not known for any particle. The nucleon 
EMT form factors are defined as \cite{Pagels}
\ba\label{Eq:definition-FF}
   \la p^\prime| \hat T_{\mu\nu}|p\rangle
   &=& \bar u^\prime\biggl[
   A(t)\,\frac{\gamma_\mu P_\nu+\gamma_\nu P_\mu}{2}+
   B(t)\,\frac{i(P_{\mu}\sigma_{\nu\rho}+P_{\nu}\sigma_{\mu\rho})\Delta^\rho}{4M_N}
   + D(t)\,\frac{\Delta_\mu\Delta_\nu-g_{\mu\nu}\Delta^2}{4M_N}
   \biggr]u\, ,\;\;\;\;\;\;\;\;\;\;
   \label{Eq-app:ff-of-EMT-alternative} \ea
with $P=(p+p')/2$, $\Delta=(p'-p)$, $t=\Delta^2$ and nucleon spinors 
$\bar u^\prime = \bar u(p^\prime)$, $u=u(p)$ normalized as 
$\bar u\, u=2 M_N$.
Other notations include $\;M_2(t)=A(t)$, $\;2\,J(t)=A(t)+B(t)$, 
$\;D(t)=4\,C(t)=\frac45\,d_1(t)$.
At zero momentum transfer the form factors satisfy the constraints
\ba\label{Eq:constraints-FF}
	A(0)=1 \, , \;\;\;
	B(0)=0 \, , \;\;\;
	D(0)=D = \mbox{unknown}.
\ea
The first relation, interpreted in infinite momentum frame, means
that nucleon's constituents carry its total momentum.
The second, known as the vanishing of the nucleon's
gravitomagnetic moment, is equivalent to $J(0)=\frac12$ which means 
the nucleon's constituents carry its total angular momentum.
The third relation in (\ref{Eq:constraints-FF}) defines what 
we mean by the $D$-term in this work.

The EMT form factors can be accessed via 2nd Mellin moments 
of unpolarized generalized parton distributions (GPDs) of quarks and
gluons \cite{Muller:1998fv,Ji:1996ek,Radyushkin:1996nd,
Collins:1996fb,Goeke:2001tz,Ji:1998pc}. 
This allows one to study the nucleon's spin decomposition \cite{Ji:1996ek}, 
and also the ``mass decomposition'' can be addressed 
\cite{Ji:1994av,mass-workshop}. This information can be accessed through the
quark and gluon contributions to $A(t)$ and $B(t)$.

But what exactly does $D(t)$ tell us about the nucleon?
It is the purpose of this proceeding to highlight the physics
associated with $D(t)$, and to discuss recent developments
and applications.

\section{$\mathbf{D}$-term, the last unknown global property}

The specific relation of the $D$-term to GPDs was clarified in 
\cite{Polyakov:1999gs}. (Notice that ``our'' $D(t)$ corresponds 
to the leading term in the Gegenbauer expansion of the
``$D$-term'' defined in \cite{Polyakov:1999gs}.) Further aspects 
were discussed in \cite{Goeke:2001tz,Teryaev:2001qm}.
Similarly to the way electric form factors provide insights on the 
electric charge distribution \cite{Sachs}, EMT form factors offer
insights on energy density $T_{00}(r)$, orbital angular momentum density, 
and stress tensor $T_{ij}(r)$ related to the $D$-term \cite{Polyakov:2002yz}. 
The interpretation 
is performed in the Breit-frame characterized by $\Delta^0=0$ such
that 3D-Fourier transforms can be performed. This procedure is known 
to suffer from relativistic corrections \cite{Ji:1998pc}, and  justified 
only if the particle is heavy such that its Compton wavelength is much 
smaller than its ``internal size.'' The nucleon is a heavy object
in the limit of a large number of colors $N_c$ in QCD. But even in 
real life are the relativistic corrections small for the nucleon,
and safely negligible for nuclei \cite{Jonathan-new}.

The $D$-terms of pions, nucleons and nuclei were investigated in a variety
of theoretical approaches, including
 free field theories,
 soft-pion theorems, 	% \cite{Polyakov:1999gs},
 liquid drop model,  	% \cite{Polyakov:2002yz}, 
 chiral perturbation theory, 
			% \cite{Donoghue:1991qv,Kubis:1999db,Megias:2004uj},
 bag model,		% \cite{Ji:1997gm}, 
 chiral quark soliton model,
 			% \cite{Petrov:1998kf,Schweitzer:2002nm,Goeke:2007fp},
 lattice QCD,		% \cite{Hagler:2003jd}, 
 nuclear models,	% \cite{Liuti:2005gi,Guzey:2005ba},
 Skyrmions, and		% in free space \cite{Cebulla:2007ei} and  
 			% embedded in nuclear medium \cite{Kim:2012ts},
 dispersion relations  	% \cite{Pasquini:2014vua},
 			% and a % chiral pion model \cite{Son:2014sna}.
\cite{Polyakov:1999gs,Polyakov:2002yz,Jonathan-new,
Novikov:1980fa,Voloshin:1980zf,Donoghue:1991qv,
Kubis:1999db,Megias:2004uj,Chen:2001pv,Ji:1997gm,
Petrov:1998kf,Schweitzer:2002nm,Goeke:2007fp,Goeke:2007fq,
Hagler:2003jd,Liuti:2005gi,Guzey:2005ba,Cebulla:2007ei,Kim:2012ts,
Pasquini:2014vua,Son:2014sna}.
Also $D$-terms of photons \cite{Gabdrakhmanov:2012aa},
$Q$-balls and $Q$-clouds and their excitations 
\cite{Mai:2012yc,Mai:2012cx,Cantara:2015sna,Bergabo:new},
and $\Delta$-resonances \cite{Perevalova:2016dln} were studied.
In all theoretical approaches the $D$-terms of various particles 
were found negative. 

To understand why the $D$-term is the ``last unknown global property'' 
of the nucleon, we recall that the structure of strongly interacting 
particles is typically probed by means of the other 
fundamental forces: electromagnetic and weak interaction and, 
at least in principle, the gravity. The particles couple to the interactions
via currents $J^\mu_{\rm em}$, $J^\mu{\!\!\!\!}_{\rm weak}$, $T^{\mu\nu}_{\rm grav}$
which are conserved (in case of weak interactions we deal with the 
partial conservation of the axial current, PCAC).
The matrix elements of these currents are described in terms of form 
factors which contain a wealth of information on the probed particle. 
The undoubtedly most fundamental information corresponding to the 
form factors at zero momentum transfer: the ``global properties'' 
electric charge $Q$, magnetic moment $\mu$, 
axial coupling constant $g_A$, induced pseudo-scalar coupling constant
$g_p$, mass $M$, spin $J$, and $D$-term $D$. 
All global properties are well-known (see table below) and can be 
looked up e.g.\ in the particle data book, except for the $D$-term.
See the following table for an overview.

\vspace{3mm}

\begin{tabular}{rllrrrl} 
		\hline
{\bf em:}	& $\partial_\mu J^{\mu^{\phantom X}}_{\rm em}=0$ 
		& $\la N^\prime|J^\mu_{{\bf em}}|N\ra$ 
		& $\longrightarrow$  
		& $Q_{\rm prot}$ 
		& $=$ & $1.602176487(40)\times10^{-19}$C\\
		&&&
		& $\mu_{\rm prot}$ 
		& $=$ & $2.792847356(23) \mu_N$ \\ \hline
{\bf weak:}	& PCAC 
		& $\la N^\prime|J^{\mu^{\phantom X}}_{{\bf weak}}|N\ra$ 
		& $\longrightarrow$  
		& $g_A$        	
		& $=$ & $1.2694(28)$ \\
		&&&
		& $g_p$ 
		& $=$ & $8.06(0.55)$
		% from MuCap, Paul-Scherer Institute, arXiv:1210.6545 [nucl-ex] 
		\\ \hline
{\bf gravity:} & $\partial_\mu T^{\mu\nu}_{{\bf grav}}=0$ 
		& $\la N^\prime|T^{\mu\nu^{\phantom X}}_{{\bf grav}}|N\ra$ 
		& $\longrightarrow$  
		& $M_{\rm prot}$		
		& $=$ & $938.272013(23)\,$MeV$/c^2$ \\
		&&&
		& $J$ 		
		& $=$ & $\frac12$\\
		&&&
		& {\boldmath{$D$}} & $=$ & {\bf ?}
\\ \hline
\end{tabular}

\section{\boldmath 3D densities of the EMT}

As shown in \cite{Polyakov:2002yz} performing the 3D Fourier-transforms 
of the form factors in Breit frame yields the components of static EMT 
$T_{\mu\nu}(\vec{r},\vec{s})$ where $\vec{s}$ denotes the nucleon 
polarization vector in its rest frame. 
$T_{00}(r)$ and $T_{0k}(\vec{r},\vec{s}\,)$ yield the 
energy and angular momentum densities which are obviously normalized: 
integrating them over space yields respectively the mass and the spin 
of the particle.
With $\vec{e}=\vec{r}/r$, $r=|\vec{r}\,|$
the stress tensor (for spin-0 or spin-$\frac12$) is given by
\be\label{Eq:T_ij-pressure-and-shear}
    T_{ij}(\vec{r}\,)
    = s(r)\left(e_ie_j-\frac 13\,\delta_{ij}\right)
        + p(r)\,\delta_{ij}\, , 
\ee
where $p(r)$, $s(r)$ denote the pressure and shear force distributions.
EMT conservation $\partial^\mu\hat{T}_{\mu\nu}=0$
implies for the static stress tensor $\nabla^i T_{ij}(\vec{r}\,)=0$.
This in turn implies that $p(r)$ and $s(r)$ satisfy
\be\label{Eq:diff-eq-s-p}
    \frac23\;\frac{\partial s(r)}{\partial r\;}+
    \frac{2s(r)}{r} + \frac{\partial p(r)}{\partial r\;} = 0\;.
\ee
It moreover implies the von Laue condition \cite{von-Laue}, 
a necessary (not sufficient) condition for stability,
\be\label{Eq:stability}
    \int\limits_0^\infty \!\di r\;r^2p(r)=0 \;.
\ee
The $D$-term can be expressed in two different ways in terms of 
shear and pressure distributions as
\ba\label{Eq:D-from-p-and-s}
        D 
	\; \stackrel{({\rm a})}{=} \; m \int\di^3 r\;r^2\, p(r)
	\; \stackrel{({\rm b})}{=} \;-\,\frac{4}{15}\,m\int\di^3r\;r^2\,s(r)\;.
\ea
Due to mechanical stability arguments we expect the densities
to comply with the constraints \cite{Perevalova:2016dln}
\be\label{Eq:local-criteria}
	\mbox{(a)} \;\;\;\; 
	T_{00}(r) \ge 0 \, , \;\; \;\; \;\;\; 
	\mbox{(b)} \;\;\;\; 
	\frac23\,s(r) + p(r) \ge 0 \, .
\ee

\section{\boldmath Klein-Gordon particles and Goldstone bosons}

The first $D$-term calculation, of a scalar boson in free 
Klein-Gordon theory, was reported in~\cite{Pagels}. Defining the two 
EMT form factors of a scalar particle analog to Eq.~(\ref{Eq:definition-FF})
one finds $D=-1$ \cite{Jonathan-new}. One finds $D=-1$ also for
the Goldstone bosons of spontaneous chiral symmetry breaking in 
the soft-pion limit \cite{Novikov:1980fa,Voloshin:1980zf}, which can also 
be derived from a soft-pion theorem for pion GPDs \cite{Polyakov:1999gs}. 
Chiral corrections up to ${\cal O}(E^4)$ (with $E$ denoting masses 
or momenta) to EMT form factors were computed for 
real world pions, kaons and $\eta$-mesons in chiral perturbation theory 
\cite{Donoghue:1991qv,Kubis:1999db}.
It is customary to assign an ``internal size'' to these hadrons 
through the electric form factor of e.g.\ $\pi^+$ according to
$F(t)=1+\frac16t\la r^2\ra_\pi+{\cal O}(t^2)$. However, pions, kaons and
even $\eta$-mesons are too light to discuss 3D densities \cite{Jonathan-new}. 
The situation is different for nuclei \cite{Jonathan-new}, 
which we shall discuss next.

%===== BEGIN FIGURE 1: NUCLEUS LIQUID DROP =======================
\begin{wrapfigure}{HR}{5cm}

\vspace{-5mm}

\centering
\includegraphics[width=4.5cm]{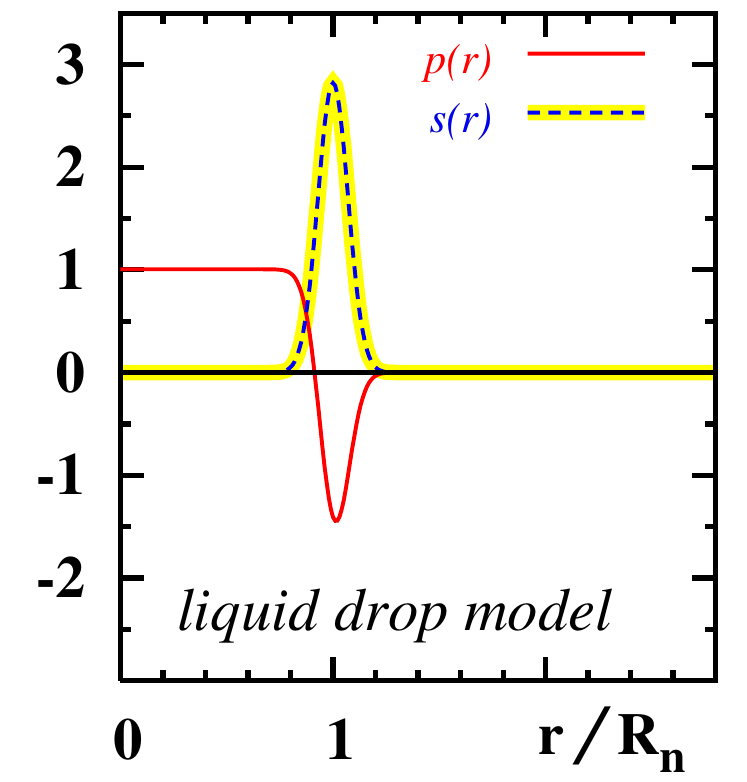}
\caption{\label{FIG-01:liquid-drop}
	The pressure and shear forces of nuclei 
	(in units of $p_0$) as functions of $r$ (in units of 
	nuclear radius $R_n$) in the liquid drop model. }
\end{wrapfigure}
%======= END FIGURE 1 ============================================

\section{\boldmath Nuclei}

In \cite{Polyakov:2002yz} also first insights into the physics of 
$D$-terms of nuclei were given: assuming nuclei (in the liquid drop model)
to have ``sharp edges'' pressure and shear forces are given by
$p(r) = p_0\,\theta(R_n-r)-\frac13\;p_0R_n\,\delta(R_n-r)$ and
$s(r) = \gamma\,\delta(R_n-r)$ with surface tension
$\gamma = \frac12\;p_0R_n$. This yields for the $D$-term
\be\label{Eq:Dterm-nucleus}
	D_{\rm nucleus} =-\frac45\,\biggl(\frac{4\pi}{3}\biggr)
	\,m_n\gamma\, R_n^4 \;.
\ee
% The $D$-term is connected to the interaction (surface tension) 
% binding the nucleus. 
Finite skin-effects make the 
$\Theta$- and $\delta$-functions in $p(r)$ and $s(r)$ smooth, 
see Fig.~\ref{FIG-01:liquid-drop}, and the $D$-term more negative. 
Remarkably $D_{\rm nucleus}\propto A^{7/3}$ since the nuclear masses and radii
grow like $m_n\propto A$ and $R_n\propto A^{1/3}$ with the mass number
\cite{Polyakov:2002yz}. Calculations in more sophisticated nuclear
models support this prediction \cite{Guzey:2005ba}.

\section{Nucleon}

\begin{figure}[t!]
\centering
\includegraphics[height=4.1cm]{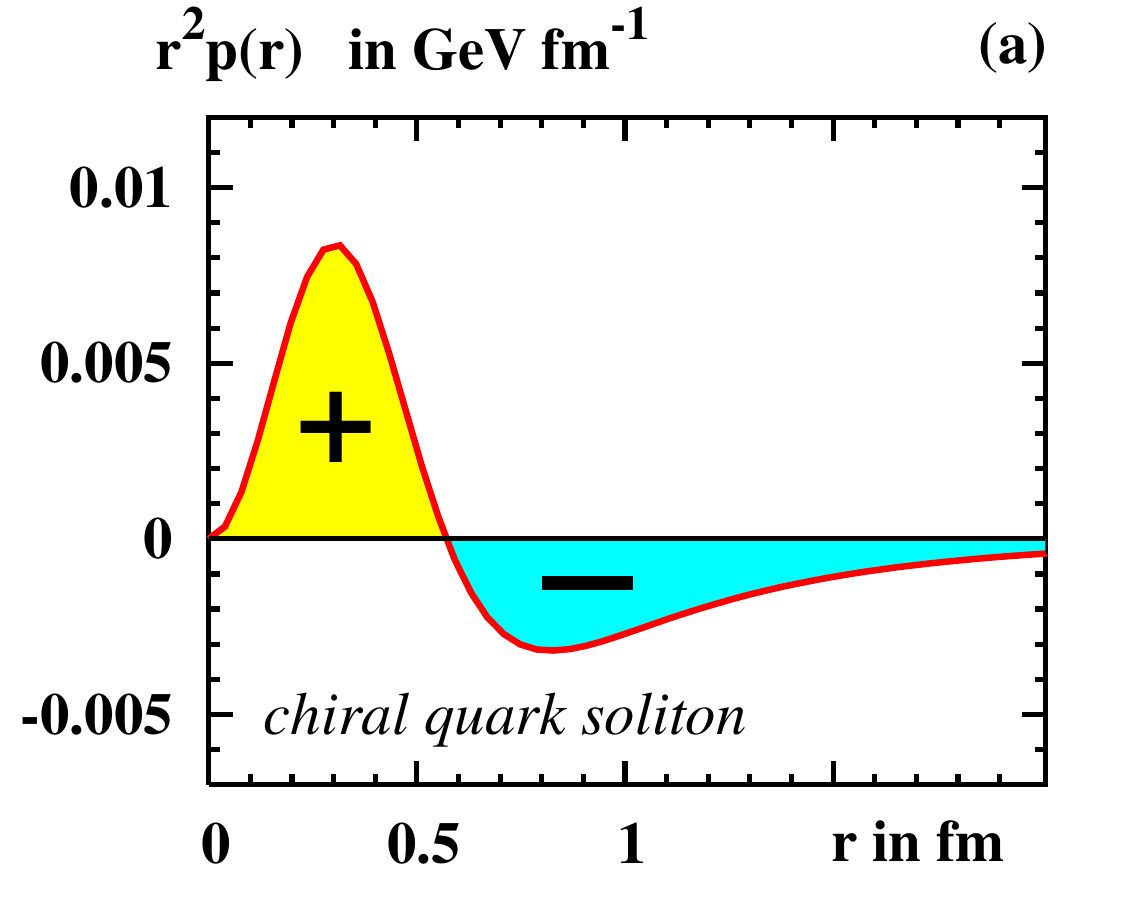}%
\includegraphics[height=4.1cm]{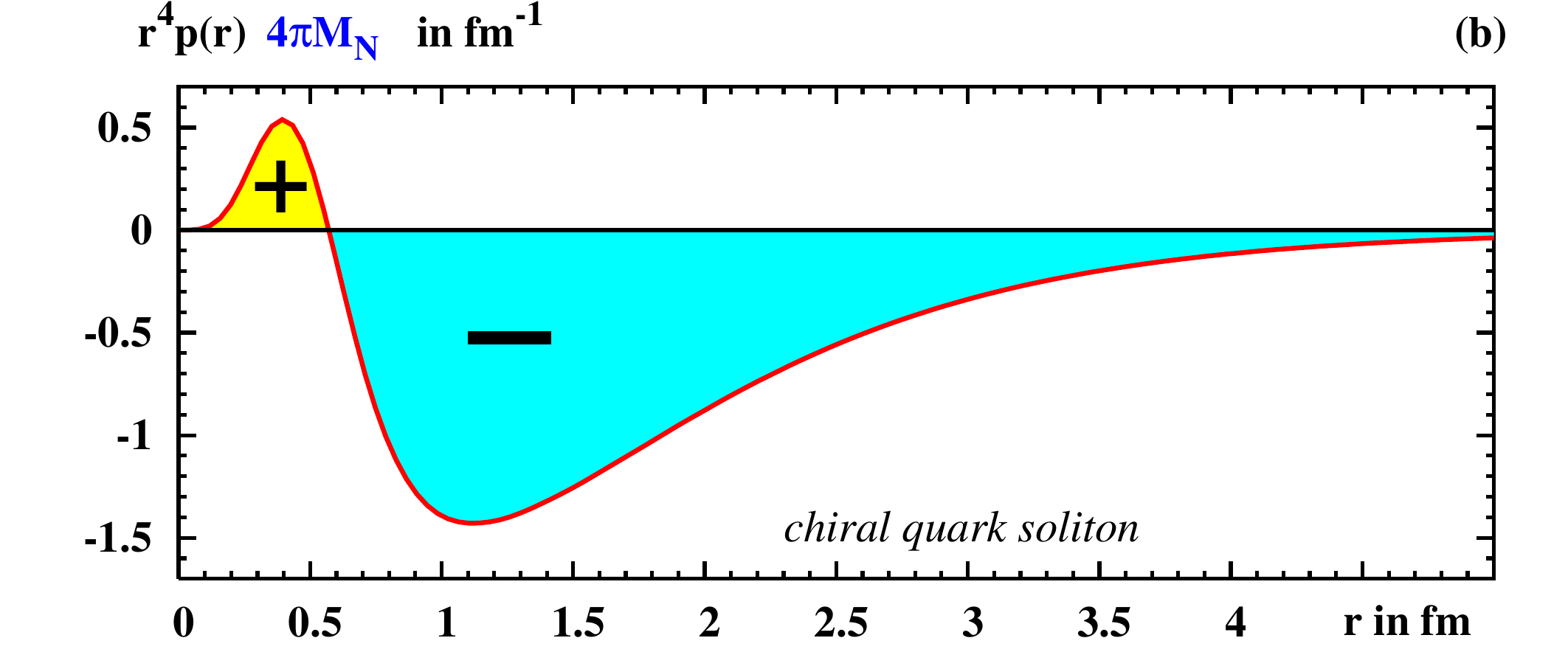}
\caption{\label{FIG-02:pressure-CQSM} 
	The pressure distribution inside nucleon from 
        chiral quark soliton model \cite{Goeke:2007fp}.
	(a) the shaded areas of $r^2p(r)$ above and below $y$-axis are
	exactly equal to each other demonstrating how the von Laue condition 
	% (\ref{Eq:stability}) 
	is realized.
	(b) $r^4p(r)$ is (up to a prefactor) the integrand of 
	the $D$-term demonstrating why $D<0$.}
\end{figure}

Also the $D$-term of the nucleon is negative \cite{Petrov:1998kf}.
Fig.~\ref{FIG-02:pressure-CQSM} shows predictions for the pressure
from the chiral quark-soliton model \cite{Goeke:2007fp}: $p(r)$ is
positive for $r<r_0$, and negative for $r>r_0$ with 
$r_0\simeq\;$(0.5--0.6)$\,$fm.
Positive $p(r)$ in the inner region means repulsion.
This is intuitively interpreted as the ``Fermi pressure'' due to 
``Pauli blocking'' in the quark core. Negative pressure in the outer 
region means attraction which is associated with the chiral soliton fields
% (in the model \cite{Goeke:2007fp} one may speak loosely of the pion cloud) 
responsible for binding the quarks. The repulsive and attractive forces 
balance each other {\it exactly} according to the von Laue condition 
(\ref{Eq:stability}), as shown in Fig.~\ref{FIG-02:pressure-CQSM}a. 
This multiplied by $r^2$ (and by the prefactor $4\pi M_N$ with  
nucleon mass $M_N$) yields the integrand of $D$
in Eq.~(\ref{Eq:D-from-p-and-s}a) depicted in 
Fig.~\ref{FIG-02:pressure-CQSM}b which ultimately yields a negative $D$.
In early works it was therefore conjectured that the negative sign
of the $D$-term is a consequence of stability \cite{Goeke:2007fp}.
This is not incorrect, but not the full story as we shall see later. 
In the chiral limit $p(r)$ and $s(r)$ behave like $1/r^6$ at large distances. 
Therefore $D$ exists in the chiral limit, but $D^\prime(t)|_{t=0} \propto 1/m_\pi$ 
diverges \cite{Goeke:2007fp}. 
The predictions from the chiral quark soliton model were shown 
\cite{Goeke:2007fq} to be in good agreement with available lattice 
QCD results \cite{Hagler:2003jd}.

%===== BEGIN FIGURE 2: CQSM PRESSURE =============================
%======= END FIGURE 2 ============================================

\section{\boldmath Lessons from $Q$-balls and $Q$-clouds}

%===== BEGIN FIGURE 3: Q-BALLS ===================================
\begin{figure}[b!]
\centering
\includegraphics[height=3cm]{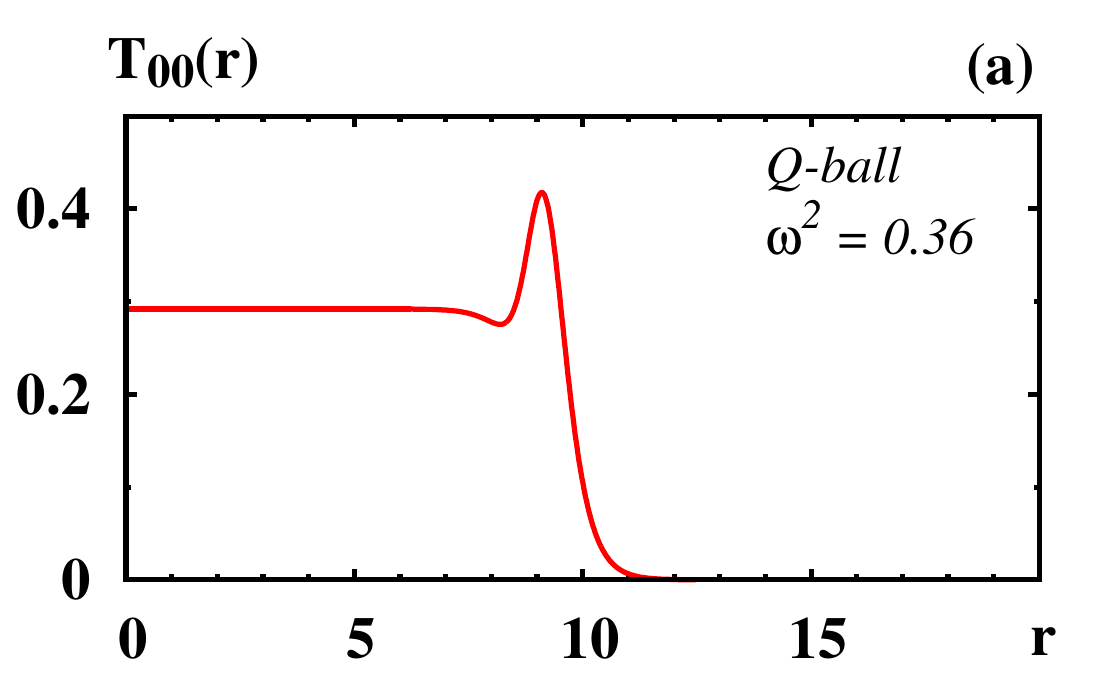} \
\includegraphics[height=3cm]{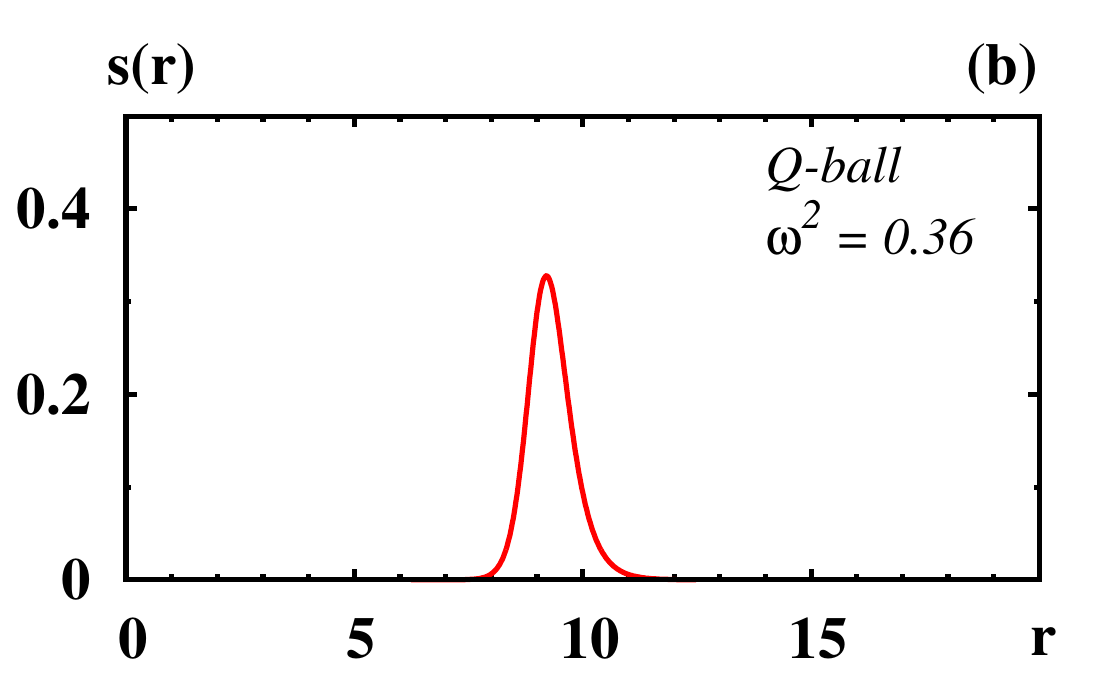} \
\includegraphics[height=3cm]{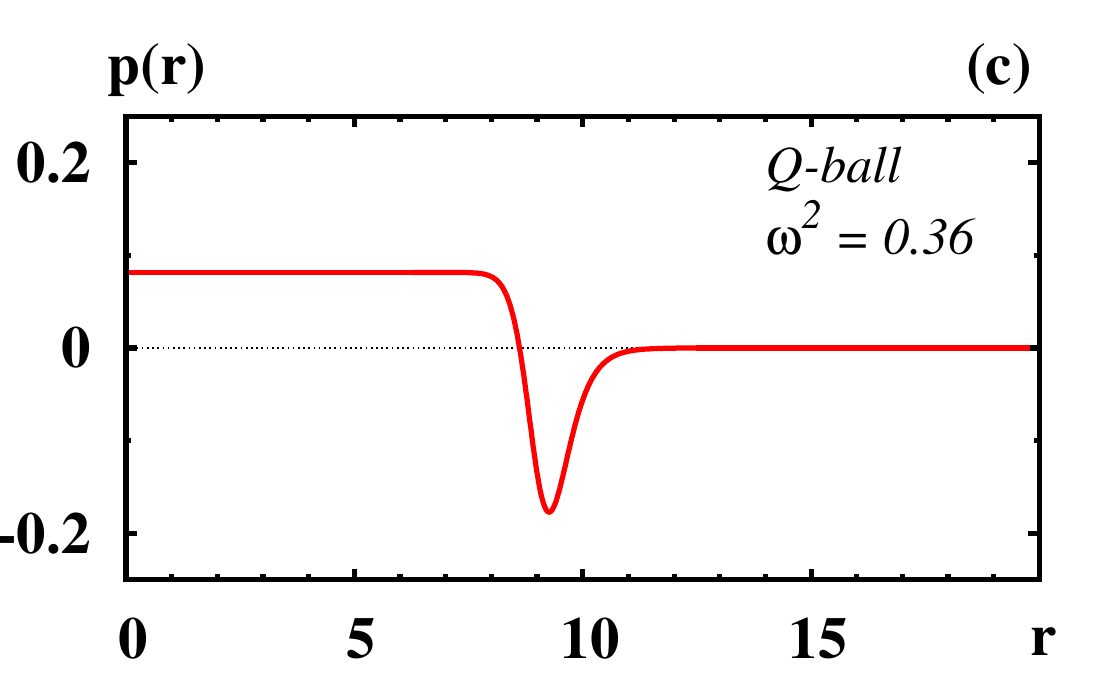}
\caption{\label{FIG-03:Q-balls}
	The energy distribution $T_{00}(r)$, shear forces $s(r)$
	and pressure $p(r)$ inside a stable $Q$-ball \cite{Mai:2012yc}.}
\end{figure}
%======= END FIGURE 3 ============================================

$Q$-balls are non-topological solitons in theories with global symmetries 
\cite{Coleman:1985ki}. They might have formed in the early universe and are
dark matter candidates. 
$Q$-balls were also used to gain insights on the $D$-term in scalar 
theories \cite{Mai:2012yc,Mai:2012cx,Cantara:2015sna,Bergabo:new}
where solitons are of the type $\Phi(t,\vec{r}\,)=e^{i\omega t}\phi(r)$ for 
$\omega_{\rm min}<\omega<\omega_{\rm max}$ with the limits determined 
by details of the theory \cite{Coleman:1985ki}.

For $\omega\to\omega_{\rm min}$ one deals with $Q$-balls \cite{Coleman:1985ki}, 
see Fig.~\ref{FIG-03:Q-balls}, absolutely stable solutions which
share features of liquid drops and have $D<0$ \cite{Mai:2012yc}. 
Interestingly, meta-stable and unstable $Q$-balls \cite{Mai:2012yc}
and their excitations \cite{Mai:2012cx} have also $D<0$.
It is therefore fair to say that stability implies a negative $D$-term.
But the reverse does need not be true: a negative $D$ does not necessarily 
imply stability.

For $\omega\to\omega_{\rm max}$ one deals with $Q$-clouds
\cite{Alford:1987vs}, unstable solutions which delocalize, 
spread out over all space, and form an infinitely 
dilute system of free quanta. 
Defining $\varepsilon^2=\omega_{\rm max}^2-\omega^2$ one finds that
at fixed $r$ energy density $T_{00}(r)$ is diluted as $\varepsilon^2$. 
At the same time $\phi(r)\propto e^{-\varepsilon\,r}$ at large $r$, and
the spatial size of the solutions diverges as $1/\varepsilon$. 
As a consequence mass and $D$-term diverge as $M\sim1/\varepsilon$ 
and $D\sim 1/\varepsilon^2$. Could this so extremely unstable 
and singular system have positive $D$? The answer is no! 
By carefully taming the divergences one finds that quantities 
like $\varepsilon M$ and $\varepsilon^2D$ have
well-defined limits for $\varepsilon\to0$, and the $D$-term
expressed in these units is negative \cite{Cantara:2015sna}.

If even such extreme instabilities as the $Q$-cloud have negative
$D$-terms, the question emerges whether it is possible to encounter 
a {\it consistent} system, however unstable it may be, which exhibits a
positive $D$-term. 
No physical system with a positive $D$-term is known so far, except
for an artifact in rigid rotator approach to the description of
baryons in large $N_c$ limit discussed in the next section.

\section{$\Delta$-resonance and unobserved states of the rigid rotator}

\begin{figure}[b!]
\centering
\includegraphics[height=3.9cm]{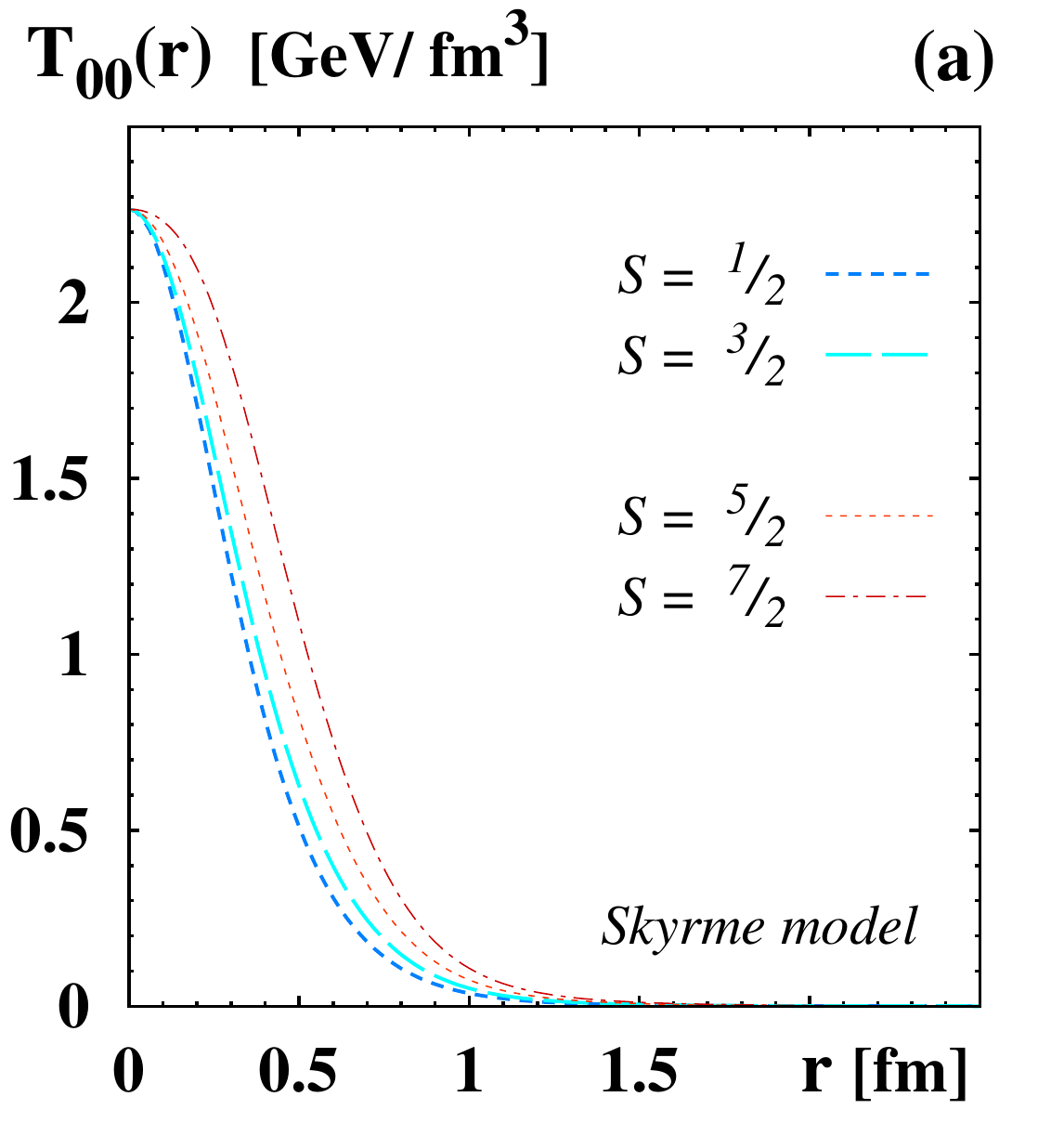}%
\includegraphics[height=3.9cm]{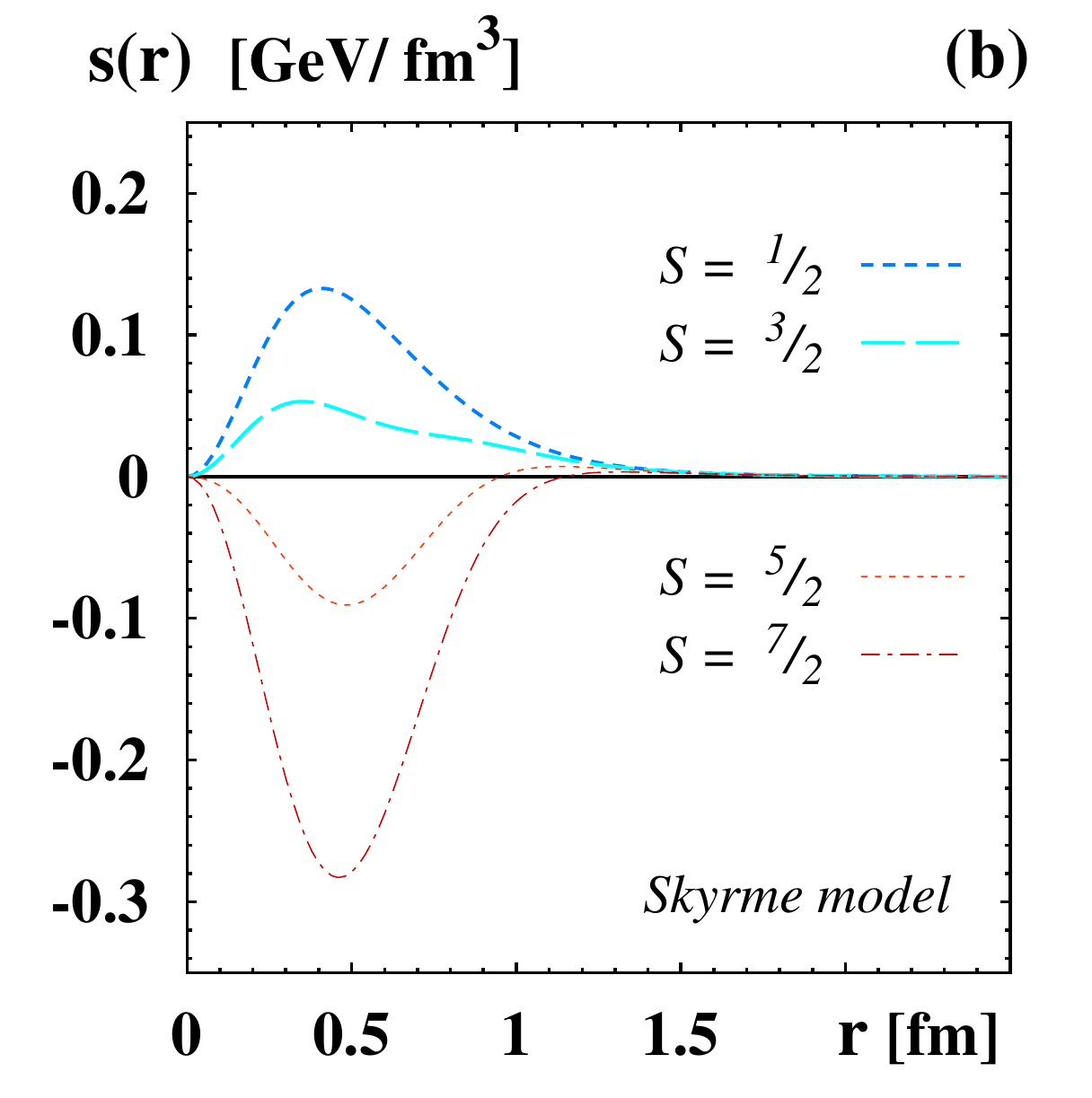}%
\includegraphics[height=3.9cm]{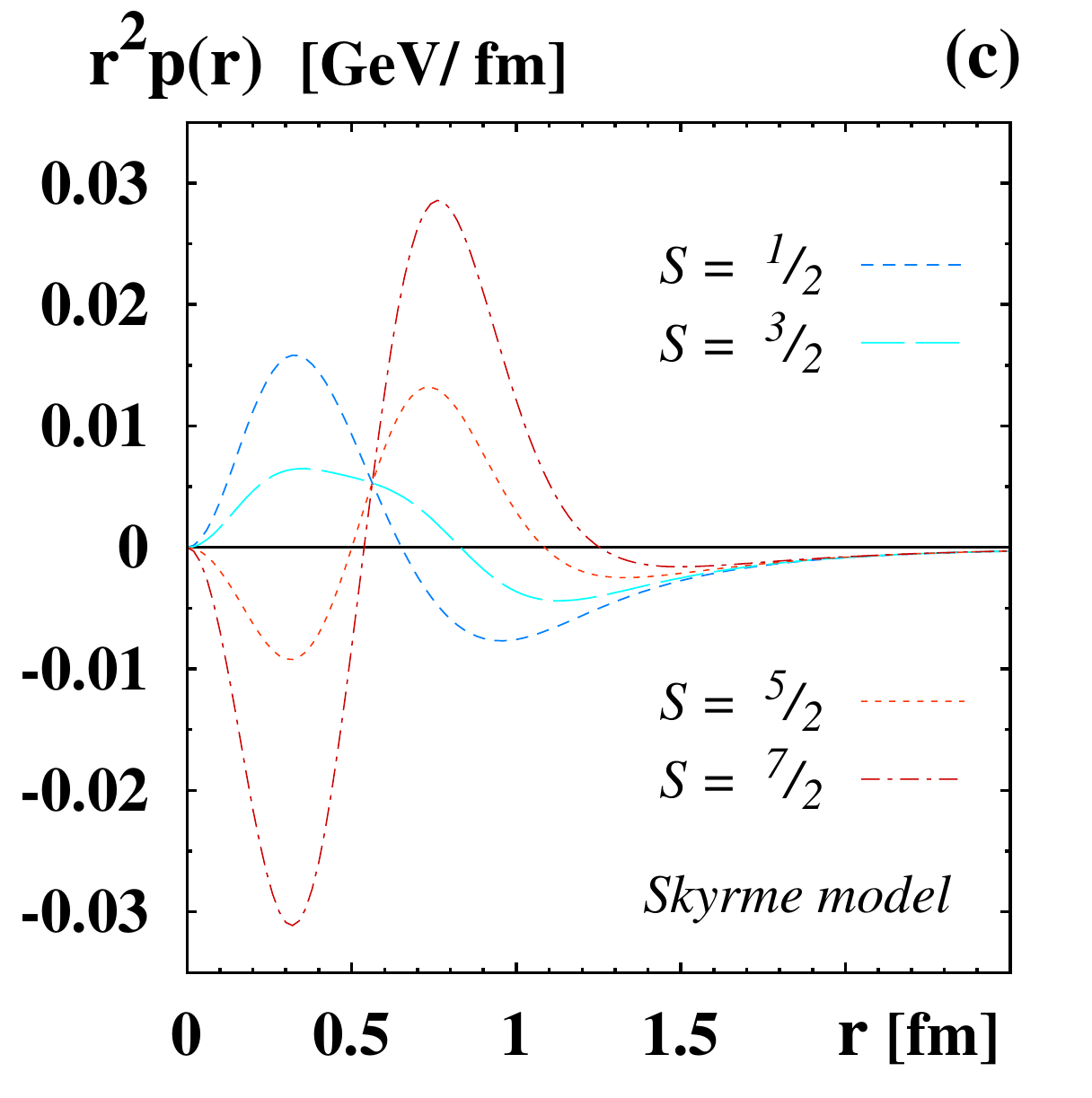}%
\includegraphics[height=3.9cm]{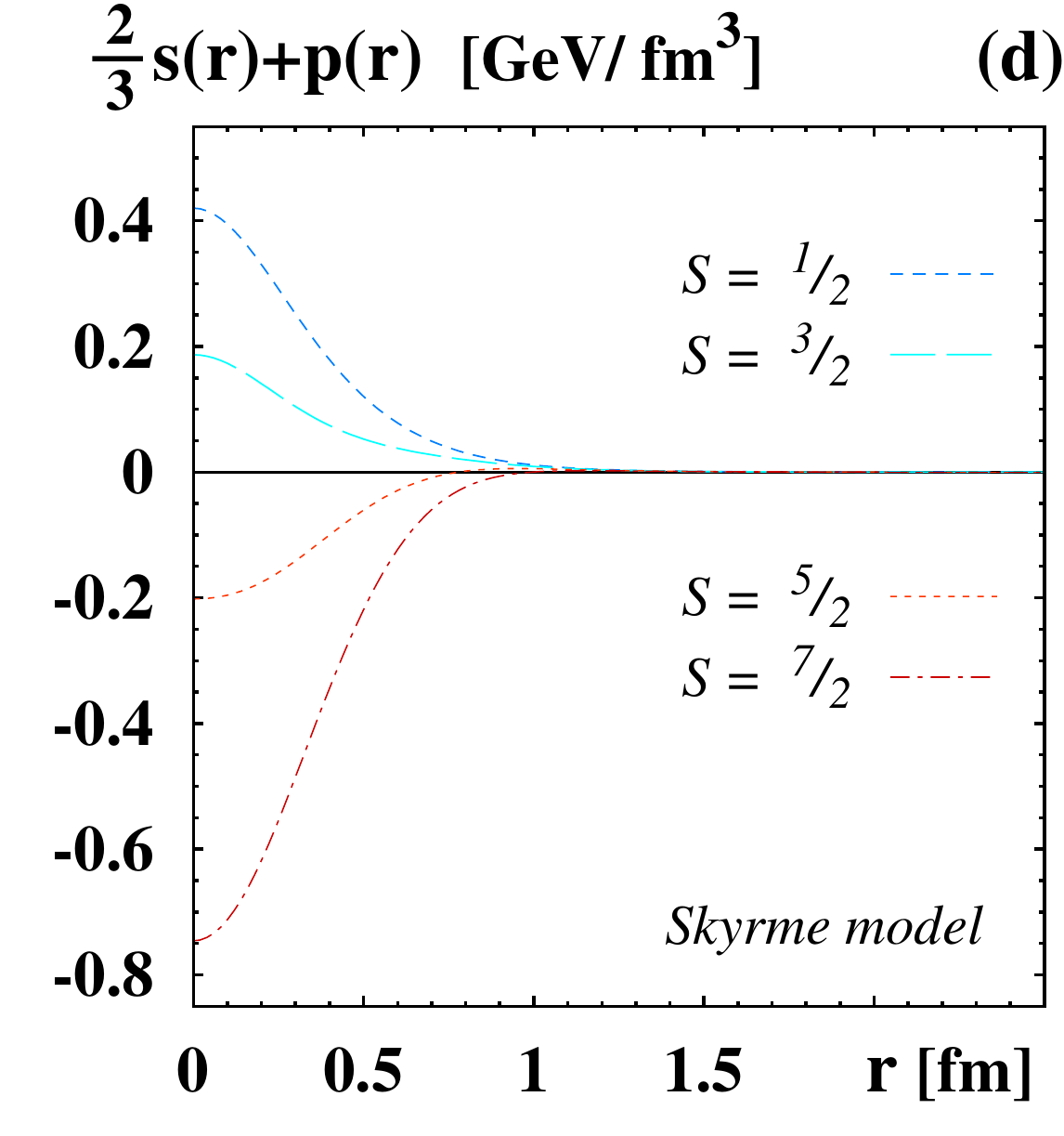}
\caption{\label{FIG-04:rigid-rotator}
	EMT densities in Skyrme model for the quantum numbers
	$S=I=\frac12,\;\frac32,\;\frac52,\;\frac72$.
	Energy density $T_{00}(r)$ indicates the states are heavier 
	with increasing spin, and does not reveal anything unusual.
	More insightful are $s(r)$ and $p(r)$  which flip 
	signs and violated the stability criterion 
	$\frac23s(r)+p(r)\ge 0$ % in Eq.~(\ref{Eq:local-criteria}b) 
	for $S=I\ge\frac52$. This calculation explains why 
	$S=I=\frac12,\;\frac32$ are physical states corresponding to nucleon 
	and $\Delta$. In contract to this $S=I\ge\frac52$ are unphysical 
	artifacts of the rigid rotator approach which are not observed in 
	nature \cite{Perevalova:2016dln}.}
\end{figure}

A similar situation is encountered with $\Delta$. It is an unstable
resonance, but nevertheless it has a negative $D$-term as shown
recently in a Skyrme model study \cite{Perevalova:2016dln}. 
Soliton models based on the large-$N_c$ expansion, 
such as the chiral quark-soliton model or Skyrme model,
describe light baryons with spin and isospin quantum numbers 
$S=I=\frac12,\;\frac32,\;\frac52,\;\dots\,$ as different rotational 
states of the same soliton solution. In this ``rigid rotator approach''
one considers $1/N_c$ corrections by expanding the action in terms of the
slow angular velocity of rotating solitons.
In contrast to the quantum numbers $S=I=\frac12$ and $\frac32$
corresponding to the nucleon and $\Delta$, one does not observe
states with $S=I\ge\frac52$ in nature. For a long time this 
unsatisfactory artifact of the rigid rotator approach was not understood.
The issue was clarified in \cite{Perevalova:2016dln}. 
Working in the Skyrme model it was shown that the $1/N_c$ corrections 
constitute a small perturbation in the nucleon case $S=I=\frac12$. 
The corrections are more sizable for $S=I=\frac32$ but one still obtains
a consistent description of the $\Delta$ with $D<0$. 
However, for faster rotating solitons $S=I\ge\frac52$ the $1/N_c$ 
corrections become so destabilizing, that the basic stability criterion 
(\ref{Eq:local-criteria}b) is violated. The results are shown in
Fig.~\ref{FIG-04:rigid-rotator}.

The energy density does not hint at anything unusual, see
Fig.~\ref{FIG-04:rigid-rotator}a: the states become heavier
with increasing spin which makes sense. The shear forces are
more insightful. All systems studied so far exhibit $s(r)\ge 0$,
and this is also the case for $S=I=\frac12,\;\frac32$ whereby $s(r)$
becomes broader for $S=\frac32$ indicating that $\Delta$ is even more
diffuse than the nucleon. But for $S=I\ge\frac52$ the shear forces 
become negative, see Fig.~\ref{FIG-04:rigid-rotator}b, and one obtains
a similarly reversed picture for the pressure which is negative 
in the center and positive in the outer region, see 
Fig.~\ref{FIG-04:rigid-rotator}c. Also this has not been found in any
other previous calculation. Both features imply at once a positive $D$-term.
Even though $p(r)$ still complies with the von Laue condition
(\ref{Eq:stability}), one deals for $S=I\ge\frac52$ with an unstable 
equilibrium. The slightest disturbance of such a system makes it explode:
matter is expelled from the center and dispersed to infinity.
The stability criterion (\ref{Eq:local-criteria}b) is violated,
see Fig.~\ref{FIG-04:rigid-rotator}d.
 
This study does not provide an example for a physical system with a positive 
$D$-term. It rather shows that these solutions are unphysical, explaining why 
the quantum numbers $S=I\ge\frac52$ are indeed artifacts of the rigid rotator
approach and not observed in nature \cite{Perevalova:2016dln}.

%===== BEGIN FIGURE 4: RIGID ROTATOR =============================
%======= END FIGURE 4 ============================================

\section{Application to hidden-charm pentaquark $\mathbf{P_c(4450)}$ at LHCb}

A practical application of EMT densities was presented recently in
\cite{Eides:2015dtr}, namely to pentaquark states observed by LHCb in 
$\Lambda_b^0 \to J/\Psi\,p\,K^-$ \cite{Aaij:2015tga}. Most of the time 
one observes weak decays $\Lambda_b^0 \to J/\Psi\,\Lambda^\ast$ followed 
by strong decays $\Lambda^\ast\to p\,K^-$. 
But the $J/\Psi\,p$ spectrum contains also structures compatible with 
pentaquark $P_c^+$ $(c\bar c uud)$ resonances.
In about $(8.4\pm0.7\pm4.2)\,\%$ of the cases a broad resonance
$P_c^+(4380)$, and in about $(4.1\pm0.5\pm1.1)\,\%$ 
of the cases a narrow resonance $P_c^+(4450)$ is formed.
These observations cannot be explained in terms of $K^-p$ resonant 
or nonresonant contributions, and are compatible with analyses of 
$\Lambda_b^0 \to J/\psi \, p \, \pi^-$ decays \cite{Aaij:2016phn}.

One way to describe such states was proposed in \cite{Eides:2015dtr}: 
the narrow $P^+_c(4450)$ can be interpreted as an $s$-wave nucleon-$\psi(2S)$ 
bound state with $J^P=\frac32{ }^-$ (these quantum numbers are not 
``most preferred'' by the analysis \cite{Aaij:2015tga}, but also
compatible with the data). The binding mechanism is due to an effective 
charmonium-nucleon interaction \cite{Voloshin:1979uv} given in terms of 
the charmonium chromoelectric polarizability and nucleon EMT densities.
In Ref.~\cite{Eides:2015dtr} also a $J^P=\frac12{ }^-$ state was predicted
with nearly the same mass as $P^+_c(4450)$ modulo hyperfine splitting 
effects which are suppressed as $1/m_Q$ in heavy quark mass limit.
The broader resonance $P_c(4380)$ does not appear as a nucleon-$\psi(2S)$ 
bound state in \cite{Eides:2015dtr}.
There are also no nucleon-$J/\Psi$ bound states, as the effective interaction 
is too weak in this channel.
In the original study \cite{Eides:2015dtr} the EMT predictions were
taken from the chiral quark soliton model \cite{Goeke:2007fp}, and
the results were confirmed in \cite{Perevalova:2016dln} using 
Skyrme model \cite{Cebulla:2007ei}. The description of $P^+_c(4450)$
as a $\psi(2S)$-nucleon bound state is therefore rather robust.

%===== BEGIN FIGURE 5: PENTAQUARKS WITH C-CBAR ===================
\begin{figure}[b!]
\centering
\includegraphics[height=3cm]{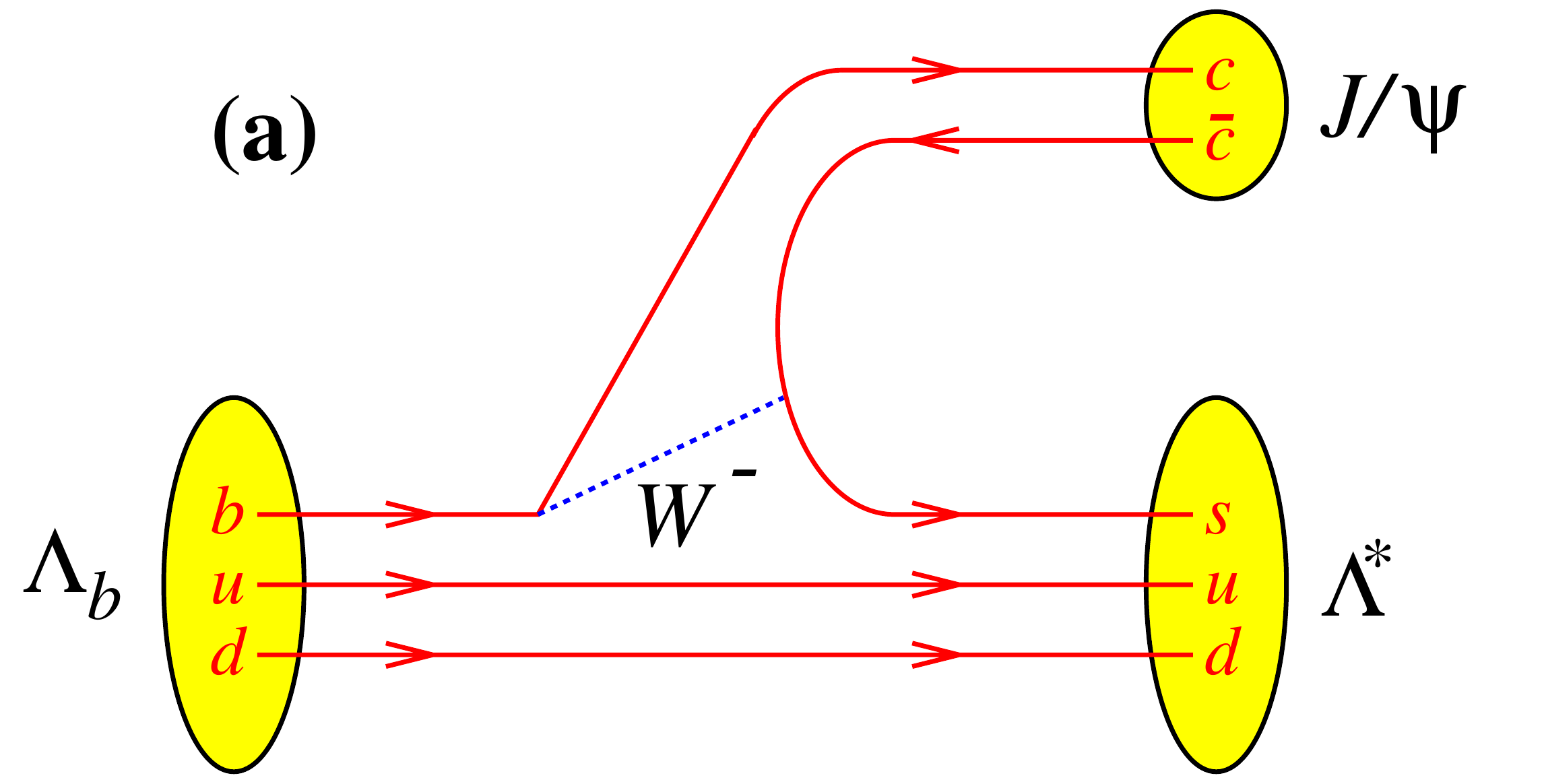}%
\includegraphics[height=3cm]{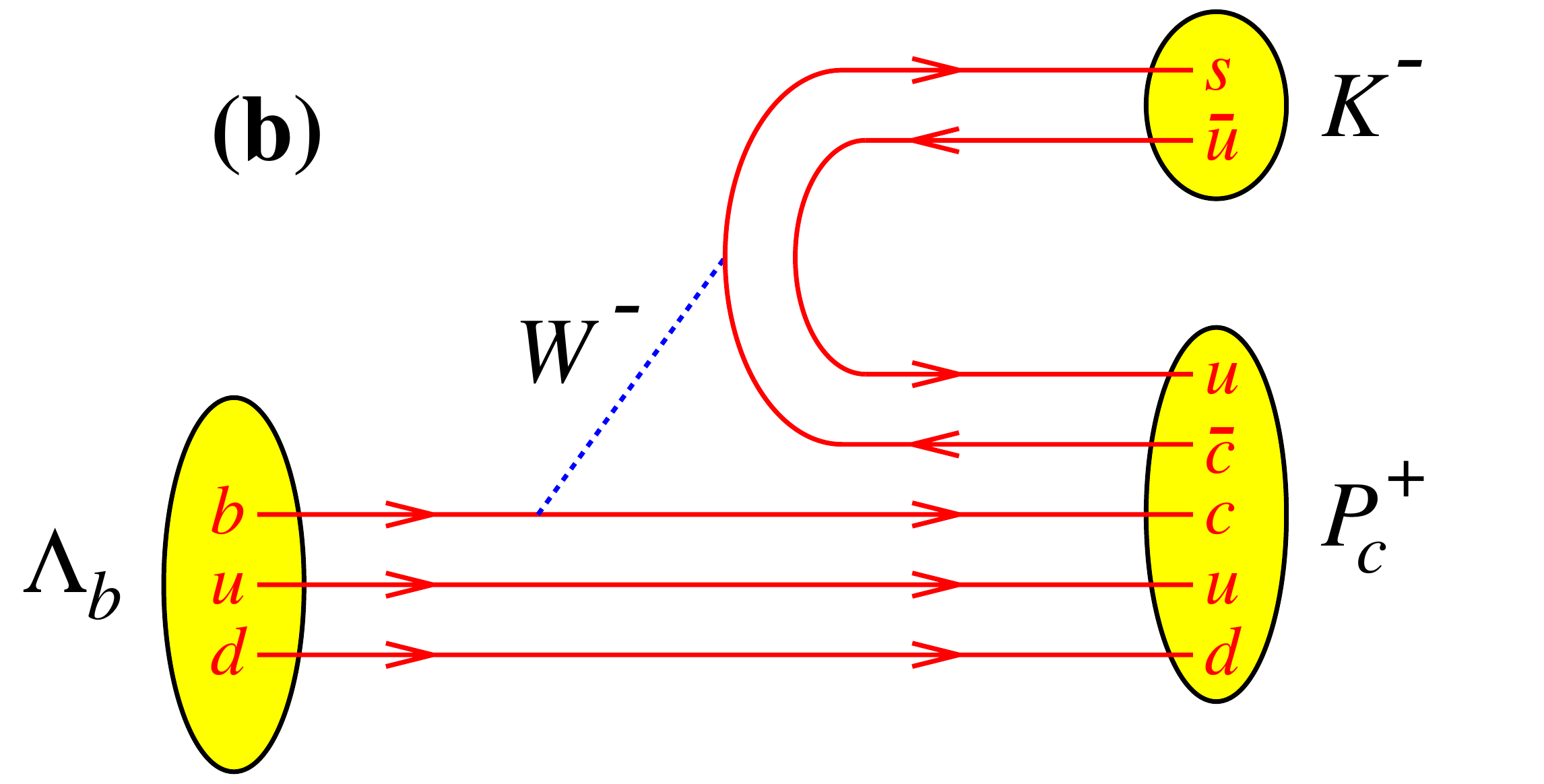}%
\includegraphics[height=3cm]{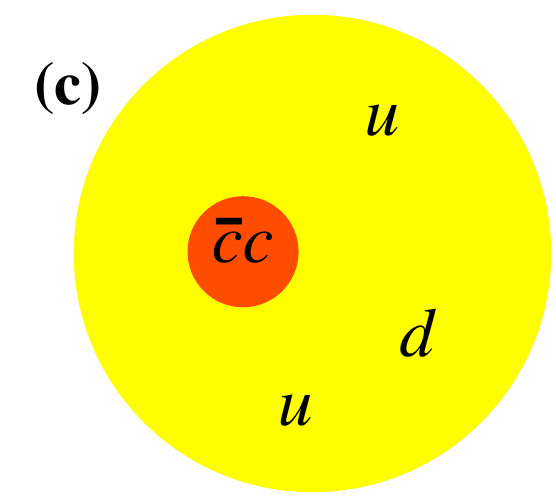}
\caption{\label{FIG-05:LHCb-pentaquarks}
	Mechanisms of $\Lambda_b^0 \to J/\Psi\,p\,K^-$ decays 
	studied at LHCb \cite{Aaij:2015tga}.
	(a) The dominant reaction is the weak decay 
	$\Lambda_b^0 \to J/\Psi\,\Lambda^\ast$ followed 
	by the strong decay $\Lambda^\ast\to p\,K^-$. 
	(b) A significant fraction of the events proceeds via the
	formation of $K^-$ and exotic resonances $P_c^+$ followed
	by subsequent $P_c^+\to J/\psi\;p$ decays.
	(c) The charmonium-baryon bound states arise because the small 
	$c\bar{c}$ penetrate the baryon, and are bound due to an
	effective potential which can be expressed in terms of EMT
	densities \cite{Eides:2015dtr,Perevalova:2016dln}.}
\end{figure}
%======= END FIGURE 5 ============================================

In  \cite{Perevalova:2016dln} also model-independent lower bounds for 
chromoelectric polarizabilities of quarkonia were derived which show
that charmonium-baryon bound states can exist for a variety of baryons.
As an application bound states in the $\Delta$-$\psi(2S)$ channel were 
predicted: a narrower negative-parity $s$-wave bound state in the around 
$4.5\,{\rm GeV}$ with width around $70\,{\rm MeV}$, and a broader 
positive-parity $p$-wave resonance around $4.9\,{\rm GeV}$ with width of 
${\cal O}(150\,{\rm MeV})$. Both actually come in a family of spin states
$J=\frac12,\,\frac32,\,\frac52$ from combining spins of $\psi(2S)$ and
$\Delta$ with small mass-splittings due to hyperfine quarkonium-baryon 
spin-spin interactions. These new states $\Pnew$  could be observed in 
weak decays of bottom-baryons $\Lambda_b^0$, $\Sigma_b$, $\Xi_b$ or
in photon-nucleon or pion-nucleon reactions.
This prediction will allow an important test of the approach. Also 
charmonium-hyperon bound states may exist \cite{Perevalova:2016dln}.

\section{Summary and Outlook}

Some aspects of the EMT structure, especially regarding the spin- or 
mass-decomposition of the nucleon, gained certain prominence in literature. 
But the prospective uses of another highly interesting direction remain largely
unexplored: the stress tensor and the $D$-term. 

Studies of the stress tensor and $D$-term in a variety of theoretical
approaches --- from soft pion theorems, to models, to lattice QCD, 
to $Q$-balls, to dispersion relations --- indicate the rich potential 
of this field. 
Among the promising new insights is the perspective to learn about the 
distribution of strong forces inside the nucleon and nuclei. 
This will shed valuable light on how the internal strong forces 
balance to form hadronic states \cite{Polyakov:2002yz}. 

Currently the EMT structure and the $D$-term of the nucleon are not 
known experimentally. It will be a long way before we will know them.
However, preliminary (and at the current stage of art necessarily 
model-dependent) projections indicate that information about the $D$-term 
and the pressure distribution of the nucleon can be mapped out through 
measurements of hard-exclusive reactions at Jefferson Lab \cite{JLab-proposal} 
or COMPASS. The experiments are in preparation.

\begin{figure}[h!]
\centering
\includegraphics[height=6cm]{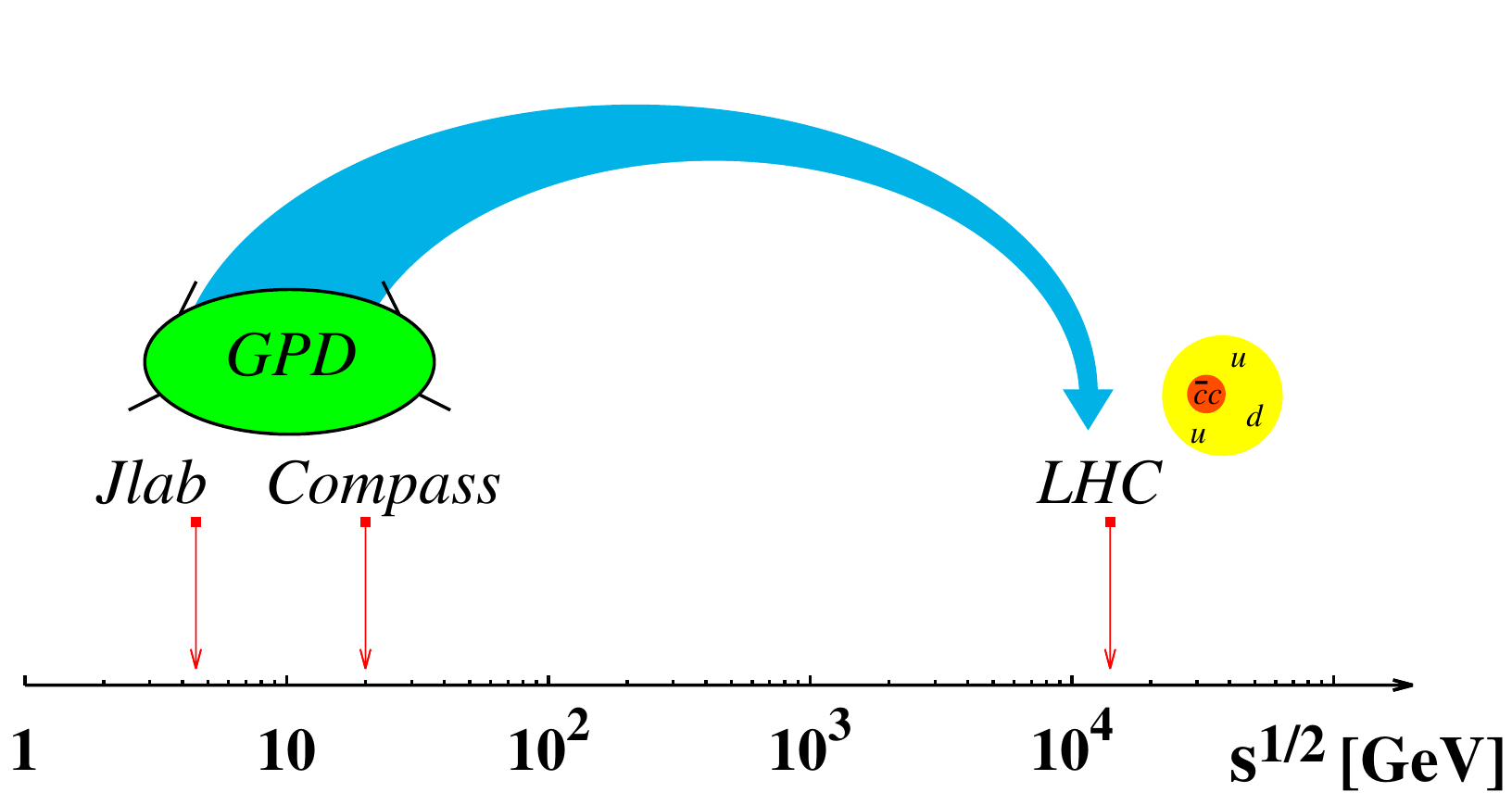}
\caption{\label{FIG-06:LHCb-pentaquarks}
	Information extracted
	from hard-exclusive processes at relatively modest
	energies via GPDs will help to constrain EMT form
	factors whose knowledge will enable us to calculate
	the effective quarkonium-baryon interaction potentially 
	allowing us to describe hidden-charm pentaquark
	states at LHCb.}
\end{figure}

An exciting development is that the knowledge of the EMT may play an
important role in the description of hidden-charm pentaquarks 
observed at LHCb \cite{Eides:2015dtr}. 
Further studies are underway and experimental tests
are needed. But the fascinating development sketched in
Fig.~\ref{FIG-06:LHCb-pentaquarks} is not unrealistic:
the knowledge gained from studies of exclusive reactions at relatively 
modest energies at Jefferson Lab and COMPASS may constrain the dynamics
of hidden-charm pentaquarks observed at LHCb. 
These exciting developments are worthwhile exploring 
and deserve further attention.

%===== BEGIN FIGURE 5: PENTAQUARKS WITH C-CBAR ===================
%======= END FIGURE 5 ============================================

\acknowledgments
P.S.\ would like to thank the organizers for creating the opportunity to
present this research, and for local support. 
This work was supported in part by the National Science Foundation 
(Contract No.~1406298), and the Deutsche Forschungsgemeinschaft
(Grant VO 1049/1).

\end{document}